\documentclass{article}
\usepackage{float}

\PassOptionsToPackage{numbers, compress}{natbib}
\usepackage[preprint]{neurips_2024}

\usepackage[utf8]{inputenc} 
\usepackage[T1]{fontenc}    
\usepackage{hyperref}       
\usepackage{url}            
\usepackage{booktabs}       
\usepackage{amsfonts}       
\usepackage{nicefrac}       
\usepackage{microtype}      
\usepackage{xcolor}         
\usepackage{amsmath}
\usepackage{graphicx}
\usepackage{wrapfig}
\usepackage{multirow}

\title{Efficient Generation of Molecular Clusters with Dual-Scale Equivariant Flow Matching}

\author{%
  \makebox[\textwidth]{%
    \begin{tabular}{ccc}
      \textbf{Akshay Subramanian} & \textbf{Shuhui Qu} & \textbf{Cheol Woo Park} \\
      \small{Massachusetts Institute of Technology} & \small{Samsung Display America Lab} & \small{Samsung Display America Lab} \\
      \texttt{akshay\_s@mit.edu} & & \\
      & & \\
      \textbf{Sulin Liu} & \textbf{Janghwan Lee} & \textbf{Rafael G{\'o}mez-Bombarelli} \\
      \small{Massachusetts Institute of Technology} & \small{Samsung Display America Lab} & \small{Massachusetts Institute of Technology} \\
      & & \texttt{rafagb@mit.edu} \\
    \end{tabular}%
  }%
}

\begin{document}

\maketitle

\begin{abstract}
Amorphous molecular solids offer a promising alternative to inorganic semiconductors, owing to their mechanical flexibility and solution processability. The packing structure of these materials plays a crucial role in determining their electronic and transport properties, which are key to enhancing the efficiency of devices like organic solar cells (OSCs). However, obtaining these optoelectronic properties computationally requires molecular dynamics (MD) simulations to generate a conformational ensemble, a process that can be computationally expensive due to the large system sizes involved. Recent advances have focused on using generative models, particularly flow-based models as Boltzmann generators, to improve the efficiency of MD sampling. In this work, we developed a dual-scale flow matching method that separates training and inference into coarse-grained and all-atom stages and enhances both the accuracy and efficiency of standard flow matching samplers. We demonstrate the effectiveness of this method on a dataset of Y6 molecular clusters obtained through MD simulations, and we benchmark its efficiency and accuracy against single-scale flow matching methods.
\end{abstract}

\section{Introduction}
Amorphous molecular solids are disordered organic systems with several applications in the field of organic optoelectronics. Their mechanical flexibility and solution processability make them attractive alternatives to inorganic semiconductors \cite{sun2022recent, li2022recent, shan2022organic}. Different molecules tend to pack differently in amorphous solids, which lead to different network extensivity, electronic properties, and transport rates \cite{kupgan2021molecular}. The influence of packing structure on optoelectronic properties has led to significant interest in computationally simulating these materials with molecular dynamics (MD) simulations \cite{kupgan2021molecular, fu2023molecular} to enhance our understanding of the correlations between structural design and corresponding packing and optoelectronic properties. This can inform better design strategies for the next generation of organic electronics devices.

Having the ability to generate amorphous solids in a computationally efficient manner can enable faster sampling from the multi-molecule Boltzmann distribution, as well as property-guided generation of optimally packed configurations. There have been several works developing efficient methods to sample from Boltzmann distributions or their emulated (non re-weighted) versions \cite{noe2019boltzmann, klein2024transferable}. Scaling up to larger systems (such as amorphous clusters) is however computationally expensive if generation is to be performed at the all-atom (AA) level. In the field of protein generation, coarse-grained (CG) representations are used to collapse amino acids into single beads, which drastically reduces system size while still retaining higher-level structural features. There have been works that generate proteins in their coarse-grained representations, and other works that back-map coarse-grained proteins into their all-atom structure \cite{kohler2023flow, charron2023navigating, yang2023chemically, jing2024alphafold}.

In this paper, we develop a unified dual-scale flow matching method that improves accuracy of all-atom flow matching methods by separating training and inference into different stages for coarse-grained and all-atom resolutions. Moreover, we show that by performing a majority of the inference compute at the coarse-grained resolution, one can reduce the time taken for inference integration without sacrificing accuracy, paving the way for more efficient generation of larger systems. As a demonstrative example, in this work, we used a dataset of amorphously packed clusters consisting of 5 Y6 \cite{yuan2019single} molecules that we obtained from MD simulations. As the coarse-grained representation for our experiments, we pre-defined a mapping scheme that collapses the 505 atoms system into 65 beads (see Figure~\ref{fig:frag_decomp}(a)). With gains in efficiency obtained from this method, we plan to scale up to larger systems in the future that are more representative of thin films in devices, as well as test other choices of coarse-graining mappings.

\section{Background}
\textbf{Conditional Flow Matching (CFM).} This is a new paradigm of training continuous normalizing flows (CNFs) efficiently in a simulation free manner \cite{albergo2023stochastic, albergo2022building, lipman2022flow}. An ODE parameterized by a learnable vector field $v_{\theta}(\mathbf{x}, t)$ transforms an easy-to-sample prior distribution $p_0$, into the more complex target distribution $p_1$. Parameters $\theta$ are learned through the regression objective
\begin{equation}
\label{CFM}
    \mathcal{L}_{CFM} = \mathbb{E}_{t \sim U[0, 1], \mathbf{z} \sim p_0(\mathbf{x_0})p_1(\mathbf{x_1}), \mathbf{x} \sim p_t(\mathbf{x} | \mathbf{z})}||v_{\theta}(\mathbf{x}, t) - u(\mathbf{x} | \mathbf{z})||^2,
\end{equation}
We chose the probability path definition to be $p_t(\mathbf{x} | \mathbf{z}) = \mathcal{N}(\mathbf{x} | t\mathbf{x_1} + (1-t)\mathbf{x_0}, \sigma^2)$ , which gives $u(\mathbf{x}|\mathbf{z}) = \mathbf{x_1} - \mathbf{x_0}$ as the conditional vector field. Similar to work by \citet{starkharmonic}, we trained $v_{\theta}$ to predict $\mathbf{x_1}$ rather than $\mathbf{x_1} - \mathbf{x_0}$, and parameterized $v_{\theta}$ by SE(3) equivariant refinement tensor field network (TFN) \cite{thomas2018tensor} layers. We present a comparison against other parameterizations in Section~\ref{VFN}.

\textbf{Harmonic Prior.} Besides the standard gaussian prior, another common choice for prior in molecule generation tasks is the harmonic prior. The harmonic prior was introduced in \citet{jing2023eigenfold} and used within the flow-matching framework by \citet{starkharmonic} as a more chemically plausible prior. The prior distribution $p_0(\mathbf{x_0})$ is interpreted as a Boltzmann distribution with a quadratic potential energy function, i.e., $p_0(\mathbf{x_0}) \propto e^{-E(\mathbf{x})} = e^{-\mathbf{x}^T\mathbf{H}\mathbf{x}}$, where $\mathbf{H}$ is chosen such that $E(\mathbf{x}) = \Sigma_{i,j \in \mathcal{E}}||\mathbf{x}_i - \mathbf{x}_j||^2$. We need to have the bond connectivity information $\mathcal{E}$ a-priori to sample from the harmonic prior.

\section{Methods}
\subsection{Model}
\textbf{Dual-Scale CFM.} This method decomposes the task of predicting all-atom coordinates $\mathbf{x_1}$, into two sequential steps of predicting coarse-grained coordinates $\mathbf{c_1} \in \mathbb{R}^{M\times3}$, and predicting all-atom coordinates $\mathbf{x_1} \in \mathbb{R}^{N\times3}$, where $M$ and $N$ are dimensions of coarse-grained and all-atom coordinates respectively. 
Two separate vector field networks (VFNs) are learned: 1) $v_{\theta}(\mathbf{c}, t)$ to generate coarse-grained bead positions $\mathbf{c_1}$, and 2) $v_{\phi}(\mathbf{x}, t | \mathbf{\hat{c}_1})$ to generate all-atom positions $\mathbf{x_1}$ conditioned on predicted bead positions $\mathbf{\hat{c}_1}$. Separate priors $p_0(\mathbf{c_0})$ and $p_0(\mathbf{x_0})$ are used for the two flows since they act on different input dimensionalities.

\textbf{Training and Inference.} Both flows are trained using Eq. \ref{CFM} with the relevant distributions and vector fields being substituted in. To be precise, the two objectives are:
\begin{equation}
\label{L_CG}
    \mathcal{L}_{CG} = \mathbb{E}_{t \sim U[0, 1], \mathbf{z} \sim p_0(\mathbf{c_0})p_1(\mathbf{c_1}), \mathbf{c} \sim p_t(\mathbf{c} | \mathbf{z})}||v_{\theta}(\mathbf{c}, t) - u(\mathbf{c} | \mathbf{z})||^2,
\end{equation}
\begin{equation}
\label{L_AA}
    \mathcal{L}_{AA} = \mathbb{E}_{t \sim U[0, 1], \mathbf{z} \sim p_0(\mathbf{x_0})p_1(\mathbf{x_1}), \mathbf{x} \sim p_t(\mathbf{x} | \mathbf{z})}||v_{\phi}(\mathbf{x}, t | \mathbf{c_1}) - u(\mathbf{x} | \mathbf{z})||^2,
\end{equation}
$v_{\phi}$ was trained using the ground-truth coarse-grained coordinates $\mathbf{c_1}$ as the condition instead of predicted coordinates $\mathbf{\hat{c}_1}$, which decoupled the two flows. Inference was performed using an Euler ODE solver to integrate from $t=0$ to $1$ for each of the two flows. For the coarse-grained flow, the bead positions were obtained with
$\mathbf{\hat{c}_1} = v_{\theta}(\mathbf{c}, t)$, and an integration timestep was performed with $\mathbf{c_{t + \Delta t}} = \mathbf{c_t} + \Delta t (\mathbf{\hat{c}_1} - \mathbf{c_0})$. Similarly, for the all-atom flow, $\mathbf{\hat{x}_1} = v_{\phi}(\mathbf{x}, t | \mathbf{\hat{c}_1})$, and $\mathbf{x_{t + \Delta t}} = \mathbf{x_t} + \Delta t (\mathbf{\hat{x}_1} - \mathbf{x_0})$ were used. A total of 40 integration steps were taken to go from coarse-grained prior sample $\mathbf{c_0}$ to all-atom target $\mathbf{x_1}$.

\begin{figure}[ht]
  \centering
\includegraphics[width=\textwidth]{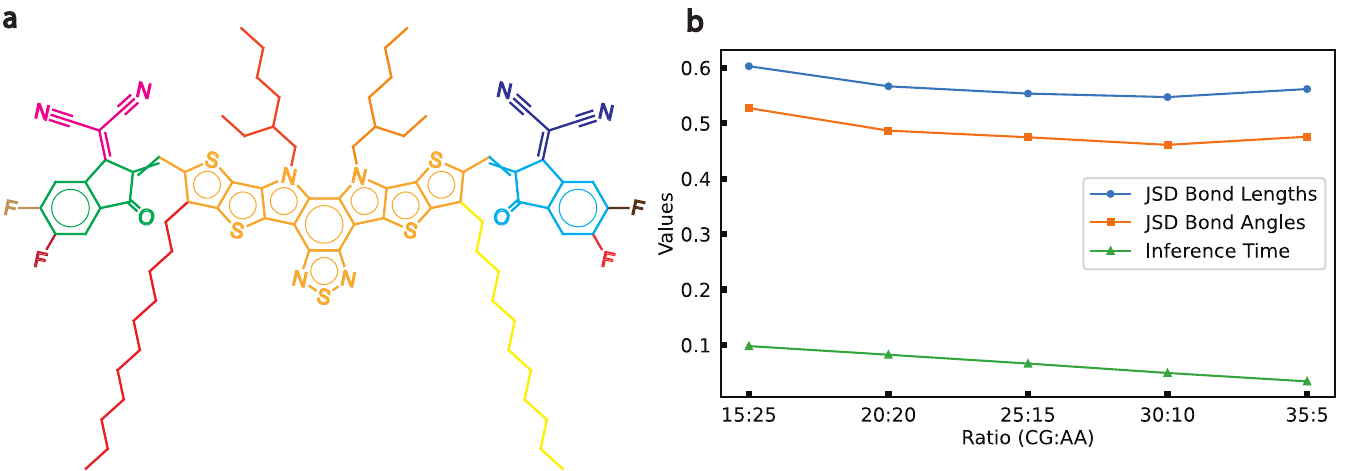}
  \caption{\textbf{Coarse-graining mapping scheme, and influence of CG:AA  ratio on metrics.} (a) Our coarse-graining mapping scheme for an individual Y6 molecule consisting of 13 beads identified with different colors. (b) JSD bond lengths, angles, and inference time on test data generation as a function of CG:AA ratio. As we increased the ratio, we observed a noticeable decreasing trend in inference time with negligible change in JSD values. }
  \label{fig:frag_decomp}
\end{figure}


\subsection{Graph Representation} 
Besides cartesian coordinates ($\mathbf{c}/\mathbf{x}$), the vector field predictors $v_{\theta/\phi}$ also take atom features $\mathbf{f^{CG/AA}}$ and bond features $\mathbf{b^{CG/AA}}$ as input. More details on features and their data-types are provided in Table~\ref{table:bond_length_angle}.

We pre-defined a mapping $B$ between node indices of all-atom atoms and coarse-grained beads, based on the basic chemical intuition that atoms present within rings have restricted degrees of freedom. The scheme is shown in Figure~\ref{fig:frag_decomp}(a). We needed to define a means of collapsing the atom features $\mathbf{f^{AA}}$ and bond features $\mathbf{b^{AA}}$ into corresponding features $\mathbf{f^{CG}}$ and $\mathbf{b^{CG}}$ of coarse-grained beads so that this information was not completely lost to the flow $v_{\theta}$. 

\textbf{Atom Features.} We aggregated atom features with $\mathbf{f^{CG}_i} = \bigoplus_{j \in B(i)}\mathbf{f^{AA}_j}$, where $\bigoplus$ denotes an arbitrary aggregation operator. In our case we used either the elementwise \texttt{Mean}  or \texttt{OR} operations depending on the data-type of the atomic feature. We did not include \texttt{String} atom features as input to $v_{\theta}$. We assigned the cartesian coordinates of beads to be the averaged coordinates of all atoms present within the bead, and included them as equivariant node features to $v_{\theta}$.

\textbf{Bond Features.} We assigned bond connectivity to beads based on connectivity of atoms within the beads. If any two atoms belonging to different beads had a bond connecting them in the all-atom structure, we created a bond between the two beads with the corresponding bond features. 

\subsection{Dataset}
We followed the approach developed in \citet{kupgan2021molecular} to simulate an amorphous morphology for the molecules. We first performed geometry optimization of the Y6 molecular structure using the method described in \cite{subramanian2023automated}. Then we used PACKMOL \cite{martinez2009packmol} to pack 5 structures in a cubic box at a low density ($\sim$ 0.1 g/cm$^3$). The OPLS-AA force field \cite{robertson2015improved} was used via the LigParGen server \cite{dodda2017ligpargen}. The system was equilibrated for 30 ns at 650 K, cooled to room temperature (300 K) at 10 K/ns, and subjected to a 30 ns production run at 300 K. All MD simulations were performed under the NPT ensemble at 1 atm with a timestep of 2 fs, using GROMACS 2023 \cite{abraham2015gromacs}. We saved 301 frames from the production run and used a random 80:10:10 split into train, validation and test datasets. We unwrapped the molecular structures to remove periodic boundary conditions (PBC) and also removed hydrogen atoms. Each resulting frame contained 505 atoms in total in the all-atom representation.

\section{Results and Discussion}
We evaluated generated samples on 3 metrics (shown in Table~\ref{table:docking_multi_ligand}). To evaluate distributional matching capabilities, we measured Jensen Shannon Divergence (JSD) between distributions of generated and MD distributions for bond lengths and angles. We evaluated efficiency through the average time taken per integration step during inference on one NVIDIA A100 GPU. We noticed that the dual-scale flow matching method improved upon a single-scale flow matching method (with Gaussian/Harmonic priors) by 15-25\% on bond length and angle JSDs, while also decreasing the inference time by $\sim$85\%.

We also tested the influence of the ratio of integration steps taken in coarse-grained and all-atom resolutions, on the 3 metrics to see how much inference time can be saved without sacrificing accuracy (shown in Figure~\ref{fig:frag_decomp}(b)). We denote the ratio of number of CG and AA integration steps by CG:AA. It is interesting to see that even for large CG:AA ratios, we significantly decreased inference time with negligible sacrifice to JSD performance.

\begin{table}[H]
\caption{\textbf{Distribution and inference time metrics.} Comparison of single-scale CFM with Gaussian and Harmonic priors, and dual-scale CFM with Gaussian prior. We used 2 distributional metrics (on bond lengths and bond angles) that quantify the Jensen Shannon divergence (JSD) between generated and MD distributions on the test dataset. Inference time is the average time taken per integration step on one NVIDIA A100 GPU.}
\label{table:docking_multi_ligand}
\begin{center}
    \begin{tabular}{lccc}
    \toprule
        Model & Bond Lengths (JSD) & Bond Angles (JSD) & Inference Time (s) \\
    \midrule
    \textsc{Single Scale}     & 0.6563 & 0.6316 & 0.2949 \\
    \textsc{Single Scale Harmonic}              & 0.6298 & 0.6066 & 0.3039 \\
    \textsc{Dual Scale 30:10 (Ours)}              & \textbf{0.5472} & \textbf{0.4610} & \textbf{0.0496} \\
    \bottomrule
    \end{tabular}
\vspace{-0.4cm}
\end{center}
\end{table}

The preliminary results are encouraging, opening up several avenues of further exploration for the future. We plan to scale to larger systems ($\sim$200 Y6 molecules) which is more representative of active layers in devices, and also test our approach on different chemistries and system sizes. Secondly, while we pre-defined a coarse-graining mapping scheme in this paper, we plan to test other schemes in the future to identify correlations between the chosen schemes and performance. Finally, we plan on expanding to other metrics that are also typically used in Boltzmann generators and conformer generation such as effective sample size (ESS), average minimum RMSD (AMR) and coverage \cite{klein2024transferable, ganea2021geomol,jing24generative}.

\section{Conclusions}
In this work, we developed a dual-scale flow matching method that improves upon single-scale flow matching-based samplers in accuracy as well as inference efficiency, as demonstrated on a dataset of amorphous Y6 clusters. Moreover, it is interesting to see that we can push the ratio of CG:AA during inference integration to large values without sacrificing the prediction accuracy, while boosting inference speed. While we tested on clusters of size 5 in this paper, we plan to expand to larger clusters in the future whose packing properties are more representative of active layers in optoelectronic devices. We also plan on testing the influence of coarse-graining mapping scheme on performance since we tested a single pre-defined scheme in this work.

\bibliography{./bibliography}


\appendix

\section{Appendix / supplemental material}

\subsection{Choice of VFN architecture}
\label{VFN}
Our choice of VFN architecture was based on a comparison of various architectures (shown in Table~\ref{table:bond_length_angle}). We observed that using TFN as the VFN achieved best JSD performance across bond lengths and angles, as well as across gaussian and harmonic prior choices.

\begin{table}[H]
\caption{\textbf{Bond length and angle distributions comparison.} JSD comparison between TFN, EGNN, and Attentive FP with Gaussian and Harmonic priors across the two collective variables (CV).}
\label{table:bond_length_angle}
\begin{center}
    \begin{tabular}{lcccc}
    \toprule
        \textbf{CV} & \textbf{Prior} & \textbf{TFN} & \textbf{EGNN} & \textbf{Attentive FP} \\
    \midrule
    \multirow{2}{*}{\textsc{Bond Length}} 
                                    & \textsc{Gaussian}     & \textbf{0.656} & 0.791 & 0.959 \\
                                    & \textsc{Harmonic}     & \textbf{0.629} & 0.781 & 0.894 \\
    \midrule
    \multirow{2}{*}{\textsc{Bond Angle}} 
                                    & \textsc{Gaussian}     & \textbf{0.631} & 0.732 & 0.908 \\
                                    & \textsc{Harmonic}     & \textbf{0.606} & 0.770 & 0.868 \\
    \bottomrule
    \end{tabular}
\vspace{-0.4cm}
\end{center}
\end{table}

\subsection{Graph Featurization}
\begin{table}[H]
\caption{\textbf{Atom and Bond Features.} Features extracted for atoms and bonds, along with their respective data types used for featurization. Same as was used by \citet{starkharmonic}.}
\label{table:atom_bond_features}
\begin{center}
    \begin{tabular}{lcc}
    \toprule
    \textbf{Feature Name} & \textbf{Datatype} \\
    \midrule
    \textsc{atomic\_num}         & Integer \\
    \textsc{chirality}           & String \\
    \textsc{degree}              & Integer \\
    \textsc{numring}             & Integer \\
    \textsc{implicit\_valence}   & Integer \\
    \textsc{formal\_charge}      & Integer \\
    \textsc{numH}                & Integer \\
    \textsc{hybridization}       & String \\
    \textsc{is\_aromatic}        & Boolean \\
    \textsc{is\_in\_ring5}       & Boolean \\
    \textsc{is\_in\_ring6}       & Boolean \\
    \midrule
    \textsc{bond\_type}          & String \\
    \textsc{bond\_stereo}        & String \\
    \textsc{is\_conjugated}      & Boolean \\
    \bottomrule
    \end{tabular}
\vspace{-0.4cm}
\end{center}
\end{table}


\end{document}